\documentstyle[preprint,aps]{revtex} 
\tightenlines
\begin{document}

\thispagestyle{empty}

\title{Electrons and Nuclei: Fundamental Interactions and Structure}

\author{E.M. Henley}

\address{Department of Physics and Institute for Nuclear Theory, 
University of Washington,\\ 
Box 351560, Seattle, WA 98195, USA}

\maketitle

\begin{abstract}
Examples are given of the usefulness of electrons in interaction with nuclei
 for probing fundamental interactions and structure. 
\end{abstract}

\section {INTRODUCTION}
Electrons offer a great tool for testing the validity of theories of fundamental
interactions (e.g., electroweak theory) and structure. The electromagnetic
interaction of the elctron with nucleons and mesons is known and is sufficently
weak that detailed structural information about nuclei and nucleons can be
obtained. The weak interaction is less well pinned down, so that tests of the
standard model can be carried out. Where the interaction is known, there is an
additional tool for obtaining structural information, such as strangeness matrix
elements in a nucleon. Nuclei are very useful targets because it is easy to
change the charge and neutron/proton ratio. I will illustrate these remarks
with several examples:

1) Beta Decay: Tests of CVC, the unitarity of the CKM matrix within the
   standard model, 2nd. class currents and time reversal invariance (TRI).

2) Double Beta Decay and neutrino masses

3) Parity Violation and tests of the standard model; measurements of the neutron
   radii of nuclei; measurements of strangeness in the nucleon; the anapole
   moment of nuclei

4) Studies of time reversal invariance

I will not be able to cover all of these subjects in detail and will
concentrate on the more recent ones, but remind you of some of the others.

\section {BETA DECAY}

\subsection{{The Conserved Vector Current}\protect\cite{IST}}

Nuclear $\beta$ decay has been used to tests the validity of the conserved 
vector current hypothesis. The first tests, accurate to about 10\% were a 
comparison of 
the weak magnetism in beta decays of allowed isospin 1 partners $^{12} B$ and 
$^{12} N$ with the gamma decay of the same multiplet in $^{12} C$ to the ground 
state of $^{12} C$. More recent tests use polarization 
measurements, a comparison of spectra,
and a measurement of slopes to improve the accuracy of this CVC test
to a few \%. The analysis
must assume the non-existence of 2nd class currents.

What are 2nd class currents? They are weak currents with the opposite 
G-parity than the normal ones 
($G = C \exp^{-i \pi I_2}$, where C is the charge conjugation
operator and $I_2$ is the second component of the isospin operator).
An example is a term
$\bar{u}(p')\sigma_{\mu\nu} q^{\nu} \gamma_{5} u(p)$, where $q= p-p'$.
Such currents are not expected to arise at the fundamental level of the standard
  model, but can come from radiative corrections and from $m_{d} > m_{u}$. 
They have been sought assiduously, but have not yet been found; however at a 
recent conference Minamisono et al.\cite{TM} reported a preliminary sighting of 
such currents in the beta decays of mass 12 nuclei at a level of a few \%. 
It would be interesting to compare any findings with theoretical work as a test
of the standard electroweak theory; to my knowledge, no such detailed 
calculations have yet been done. 

Another use of electrons emitted in superallowed Fermi beta decays 
($0^{+} \longrightarrow 0^{+}$) in an isomultiplet is as a test of the 
standard model via unitarity of the CKM matrix.$^{1}$  Could 
there be a missing interaction, f.i.? The matrix element $V_{ud}$ connecting up 
and down quarks is by far the largest one in the unitarity of
$$ U\equiv \mid V_{ud} \mid ^{2} + \mid V_{us} \mid ^{2} + 
\mid V_{ub} \mid ^{2}  = 1 $$
    
\noindent Here $V_{ud}$ and $V_{ub}$ connect the up quark with the strange 
and bottom quarks, respectively. The precise measurements of superallowed 
transitions together with radiative corrections and removal of charge 
dependent nuclear effects allow one to determine $V_{ud}$ to better than 
$10^{-3}$. In addition, these measurements, (Fig.1), show that CVC holds to 
$\sim 4 \times 10^{-4}$. 

\vspace{2.6in}

\noindent Fig.1. Ft values for $0^+ \rightarrow 0^+ \beta$--decays.
\smallskip

\noindent A straightforward analysis of the experiments, including a recent 
$^{10}C$ experiment\cite{SJF}, gives $V_{ud} = .9740\pm.0006$. Together with 
the measurements of $V_{us}$ and $V_{ub}$, one then obtains 
$U= .9972\pm.0019$. However, it has been proposed that the nuclear 
charge--dependent corrections should be corrected by a smooth 
Z-dependence.\cite{DHW} In that case $U = .9980\pm .0019$. Is the 
discrepancy from unitarity meaningful? Personally, I doubt it. 
The largest uncertainty may reside in the charge--dependent nuclear 
corrections and there is now an attempt to push the
calculation of these corrections to increased accuracy.\cite{WCH}

Lastly let me mention tests of time reversal invariance.\cite{FB} 
I will not describe the standard searches for TRI--odd terms 
($D \vec{J}\cdot \vec{p}_{e} \times \vec{p}_{\nu}$) in beta decays, 
particularly $^{19}Ne$, where $\vec{J}$ is the spin direction of the parent 
nucleus. Recently there have been searches for the R--term, 
$R \vec{J} \cdot \vec{j} \times \vec{p}_e $ in the beta decays of 
$^{8} B$ and $^{8} Li$, where $\vec{j}$ is the spin direction of the electron; 
the advantage of these  nuclei is that they decay to the unstable $^{8}Be$, 
which breaks up into two alpha particles, which readily can  be detected. 
This experiment is underway.\cite{LDB} Another method has been used at the 
Sherrer Institute by Allet et al;\cite{MA} they measure the transverse 
polarization of the emitted electrons directly and thus find a limit 
$Im C_{T}/ C_{A}  < .012$ at the 68$\%$ level; here $C_T$ and $C_A$ are the 
coefficients of the axial tensor and axial vector currents. It would be 
useful if these experiments could be increased in accuracy so as to
at least reach the level where final state interactions spoil TRI tests; 
here this would be $7 \times 10^{-4}$. The experimenters still have a 
ways to go.

\subsection{{Double Beta Decay}\protect\cite{SPR}}

The decay $(A,Z) \longrightarrow (A,Z+2)+ e^{-}+ e^{-} +\bar{\nu} 
+\bar{\nu}$
is expected in the standard model and has
been seen seen in several nuclei ($^{82}Se$,\, $^{100}Mo$,\, and $^{150}Nd$) 
with half lives
of about $10^{20} y$, consistent with the standard model.\cite{MM}  
Searches for the no neutrino decay mode, important for determining
whether $\nu$'s are massive and of the Majorana type,  have been 
continuosuly improved; they use Ge crystals; at present the lower limit on the 
half life is $3\times 10^{24} y$ for $^{74}Ge$. 
This experiment sets a lower bound on the mass
of the most massive Majorana neutrino mass, $ m_{\nu} \sqrt{\frac{10^{24} y}
{t_{1/2}(^{76}Ge)}}$. Improved experiments
hope to push this mass down further by one to two orders of magnitude 
or find the double beta decay.

\subsection{{Parity Nonconservation Studies}\protect\cite{EMH}}

The first test of the electroweak theory and its neutral currents came from 
electron scattering on hydrogen and deuterium at SLAC. At lower energies,
parity violating (pv) scattering of electrons on nucleons and nuclei allows 
one to determine all four weak currents of the nucleons:
the axial vector and the vector couplings to protons and neutrons. The 
experiments are difficult and so far only one experiment of polarized electrons
on $^{12}C$  has been carried out at MIT.\cite{PAS}  The results agree
with the theory at a level of 10\%. 

More recently, in an ongoing experiment (SAMPLE) at MIT, the weak interaction 
of the electron and proton is being used to investigate strangeness in
the nucleon. As in all pv experiments, it is the interference of the weak
interaction with the electromagnetic one that is being detected by searching
for a parity-odd signal such as $<\vec{j}> \cdot \vec{p} $, where $\vec{p}$ 
is the incident momentum of the electron and $<\vec{j}>$ is its polarization. 

Initial evidence for non-vanishing strangeness matrix elements in the nucleon
came from measurements of the spin structure of the proton by polarized 
electrons on polarized nuclear targets. This indicated that 
\begin{equation}
\bigtriangleup s \equiv s\uparrow - s\downarrow + \bar{s}\uparrow 
-\bar{s}\downarrow \approx 0.1-0.2.
\end{equation}  
Further indications of a non-negligible fraction of strangeness came from 
elastic neutrino scattering on protons. At the present time, the axial vector 
strangeness matrix element for the proton stands at $0.1 \pm .03$.

For elastic e$^{-}$ p scattering, the standard model gives
\begin{equation}
J_\mu^{em} = \bar{u}(p')[\gamma_\mu F_{1}^{em} + i \sigma_{\mu\nu}
\frac{q^{\nu}}{2M} F_{2}^{em}] u(p),
\end{equation}
\begin{equation}   
J_\mu^{Z}= \bar{u}(p')[\gamma_\mu F_{1}^{Z}  + i \sigma_{\mu\nu}
\frac{q^{\nu}}{2M} F_{2}^{Z} + \gamma_\mu F_{A}^{Z} \gamma_{5}] u(p),
\end{equation}
If strangeness is included, then SU(3) notation is helpful, and we have
\begin{equation}
F_{1}^{em} = \frac{1}{2} \left[F_{1}^{(8)} + \tau_{3} F_{1}^{(3)}\right]
 \Longrightarrow 1  \;  {\rm for} \:p,
\end{equation}
\begin{equation}
F_{2}^{em} = \frac{1}{2}\left[(\kappa_{p} + \kappa_{n})F_{2}^{(8)} + \tau_{3}
(\kappa_{p} - \kappa_{n})F_{2}^{(3)}\right]\Longrightarrow \kappa_{p}\; 
{\rm for} \:p,
\end{equation}

\begin{equation}
F_{1}^{Z}= \frac{1}{2}\left[ -F_{1}^{(0)} + \tau_{3} y F_{1}^{(3)}
+ y F_{1}^{(8)}\right] 
\Longrightarrow \frac{1}{2} \left(1- 4\sin^{2}\theta_{W}\right) \;
{\rm for} \:p,
\end{equation}

\begin{eqnarray}
F_{2}^{Z}= \frac{1}{2}[ - g_{2} F_{2}^{(0)} + \tau_{3} y (\kappa_{p} -
\kappa_{n})F_{2}^{(3)} + y (\kappa_{p} + \kappa_{n}) F_{2}^{(8)}]\nonumber\\
\Longrightarrow \frac{1}{2} \left[\left(1 - 4 \sin^2 \theta_W \right) 
\kappa_p - \kappa_n -\kappa_s  \right]  \; {\rm for} \:p,
\end{eqnarray}

\begin{eqnarray}
F_{A}^{Z} = \frac{1}{2}\left[-g_{A}^{(0)} F_{A}^{(0)} + 2 g_{A} \tau_{3}
F_A^{(3)} + (6F - 2D) F_{A}^{(8)}\right]\nonumber \\
 \Longrightarrow  - \frac{1}{2} 
g_A^{(0)} + g_A +\left(3F - D \right)    \; {\rm for} \:p,
\end{eqnarray}

\noindent where form factors have been normalized to unity at squared 
momentum transfers $Q^{2}$ = 0, and the proton values are given at $Q^{2}$=0. 
We also use $y = \left(1 - 2 \sin^{2}\theta_{W}\right),\, g_{A}= 1.26,\, 
g_{2} = \kappa_{p} + \kappa_{n} + \kappa_{s},\, 6F-2D \simeq 1.1$ where 
F and D are the fractions that are odd and even, respectively under 
$SU(3)$; $\kappa_p$ ($\kappa_{n}$) is the anomalous magnetic 
moment of the proton (neutron), $\kappa_{s}$ is the strange magnetic moment, 
and the superscripts on the form factors refer to SU(3) transformation
properties. There are two new and unconstrained couplings and form factors,
$g_{2} F_{2}^{(0)}$ and $g_{A}^{(0)} F_{A}^{(0)}$, where 
$F_{i}^{s} = F_{i}^{(0)} + F_{i}^{(8)}$.

The Sample experiment at MIT has as its aim the measurement of $g_2F_2^{(0)} 
(Q^2)$ at small $Q^2$ where $F_2^{(0)} \approx 1$. The experimenters measure 
the cross section of 200 MeV electrons polarized parallel and antiparallel to 
their momenta on a H target. The asymmetry $a$ is given by
\begin{eqnarray}
a = \frac{d\sigma_R-d\sigma_L}{d\sigma_R+d\sigma_L} = 
- \frac{GQ^2}{\sqrt{2} \pi\alpha} \{ [ 2\tau\tan^2\frac{\theta}{2}(F_1^{em} 
+F_2^{em})(F_1^Z+F_2^Z)+F_1^{em}F_2^Z + F_2^{em} F_2^Z \tau ] \nonumber \\
-\frac{E+E'}{2M} \tan^2 \frac{\theta}{2} (1-4 \sin^2 \theta_W) F_A^Z 
(F_1^{em} + F_2^{em}) \} \nonumber \\
\times \{ (F_1^{em})^2 + \tau (F_2^{em})^2 + 2\tau\tan^2\frac{\theta}{2} 
(F_1^{em}+ F_2^{em})^2 \}^{-1}
\end{eqnarray}
where $E(E')$ is the initial (final) electron energy, $\theta$ is the 
scattering angle, and $\tau \equiv Q^2 /4M^2$. The last term in Eq.(9) is 
small because $(1-4\sin^2\theta_W) \approx 0.1$, and the second one is small at 
back angles, where the first term dominates.  The experiment is carried out 
for an average $Q^2\approx 0.1 GeV^2 (130^{\circ} < \theta < 170^{\circ})$. At 
these angles, the asymmetry is sensitive to $F_2^Z$ and thus allows a 
determination of $\kappa_s$.  The results to date\cite{BM} are shown in Fig.2,

\vspace{2.7in}
\noindent Fig.2. Results for the parity--violating asymmetry measured in the 
1995 and 1996 running periods. The hatched region is the asymmetry band (due 
to the axial radiative correction) for $\mu_s = G_M^s = 0$.

\vfill\eject
 
\noindent where the hatched band corresponds to $\kappa_s=0$ (it is a band due 
to uncertain radiative corrections). At $Q^2 = 0.1$ GeV$^2$, the authors find 
$\kappa_s F^s=0.23 \pm 0.37 \pm 0.15 \pm 0.19 n.m.$, where the first error is 
statistical, the second one is systematic and the third one is due to axial 
radiative corrections.  At present, 
$\kappa_s$ is consistent with zero, but its possible positive value 
is opposite to that predicted by most theorists\cite{MJM}, e.g. by using 
$N \rightarrow \Lambda K \rightarrow N$. The SAMPLE experiment will measure
the coupling of the Z-boson to the proton in the future; also work at MIT 
and at TJNAF is expected to bring the errors down to a level of $\pm .1$ n.m.
(nucleon magneton).

Other precision pv studies of the weak interactions of electrons and nuclei 
have been carried out with atoms.  Despite their being at lower momenta, where 
the pv effects are smaller, $\sim 10^{-11}$, these experiments have reached the 
incredible precision of $1/2\%$, which atomic theory has yet to equal; 
theoretical errors are at the level of $\sim 1\%$\cite{SAB}. At this level of 
precision the atomic experiments provide meaningful tests of the standard 
model. The dominant weak interaction term is $a_\mu V^\mu$ where the lower case 
$a_\mu$ is the axial current of the electron and $V^\mu$ is the vector current 
of the nucleus. This vector current is coherent over the nucleus, giving an 
effective charge
\begin{equation} 
Q_W = (1-4 \sin^2\theta_W)Z-N
\end{equation}
which is large for heavy atoms. The measurement on $Cs$, a one valence 
electron atom, at the $1/2 \%$ level, give $Q_W = -72.35 \pm 0.27_{exp}
\pm 0.54_{th}$.\cite{CSW} This results in an s-parameter\cite{JLR}
s = $-1.0 \pm (0.3)_{exp}\pm (1.0)_{th}$ , where s is one of the 
parameters which measures deviation from the standard model. For instance,
 this gives a lower limit on the mass of a second Z boson of 
$\sim 500$ GeV.\cite{JLR}

It has been proposed that pv measurements on a series of isotopes would allow 
one to obtain neutron radii for these nuclei. This is because $Q_W$ is 
primarily sensitive to the neutron distribution [see Eq.(10)] and $\sin^2 
\theta_W$ is well known from other measurements. This method
may prove to be the most 
accurate means of measuring differences of neutron and proton radii of
nuclei.\cite{SP}

The term $v_\mu A^\mu$ is much smaller than $a_\mu V^\mu$ because 
for the electron $v^\mu 
\propto (1-4\sin^2\theta_W) \sim0.1$ and only a single nucleon contributes to 
$A_\mu \propto <\vec{\sigma}>$, the nuclear spin. Thus, the asymmetry is 
reduced by $\geq 500$. The atomic measurements of this term make use of the 
hyperfine structure, which is due to the nuclear spin. This term has not yet 
been detected because it is hidden by the stronger nuclear anapole moment. 
What is this? It is a parity violating moment discovered by Zel'dovich in 
1957.\cite{IBZ} (I discovered it independently in 1973,\cite{EMH2} but did not
name it, but simply called it an axial coupling of the photon.) The anapole
exists only for virtual photons\cite{EMH2} that penetrate the nucleus. It is,
in reality, a combination of an electromagnetic interaction and a pv 
component of the nuclear wavefunction. This combination gives rise to 
a current similar to that in a winding on the surface of a doughnut; 
see Fig.3.\cite{RRL}

\vspace{2.5in}
\noindent Fig.3. Surface current $(\vec{J})$ on a doughnut, producing a 
toroidal magnetic field $(\vec{B})$ and anapole$(\vec{T})$.
\smallskip

\noindent The most general form for an axial coupling of the photon to a 
nucleon or other spin 1/2 particle is 
\begin{equation}
A_\mu= 
\bar{\psi}
\frac{\not{q} q^\mu - 
q^2 \gamma^\mu}{M^2}
\gamma_5 \psi a   ,
\end{equation}
where $a$ is the magnitude of the anapole moment. Note 
the need for the $q^2$ 
dependence to preserve gauge invariance. The first term vanishes for a
conserved vector current of the electron. Combined with the $ 1/q^2$ 
propagator, the second term is a contact one, unlike the electromagnetic 
interaction.

The anapole moment can be written as 
\begin{equation}
\vec a = - \pi \int d^3r r^2 \vec j(r) \nonumber \\
= \frac{1}{e} \frac{G_F}{\sqrt 2} (-1)^{J+1/2+\ell} 
\frac{J+\frac{1}{2}}{J(J+1)} 
\vec J \kappa_a ,
\end{equation}
where $J$ is the spin of the nucleus and $\kappa_a$ is a dimensionless 
constant which is a measure of the anapole moment. 
The $r^2$ allows $\vec a$ to be produced by surface currents; it comes from the
$q^2$ dependence of the axial coupling. The anapole is a purely nuclear term
which is measured in atoms. The effect is of order $\alpha G_F$, but it is
proportional to the square of the nuclear radius and therefore $ A^{2/3}$. In 
Cs and heavier atoms, it is larger by almost an order of magnitude  than the
asymmetry due to neutral weak currents. 

The recent measurement in atomic Cs at the 1/2$\%$ level\cite{CSW} has 
discovered the anapole moment in the $6S \rightarrow 7S$ transition; the 
experimenters find it through a difference in the hyperfine 
$F=4 \rightarrow 3$ and $F= 3 \rightarrow 4$ transitions of $0.077 \pm.011$ 
mV/cm. This gives $\kappa_a =0.364 \pm.062$. The theoretical value is 
sensitive to the pv $\pi$ - nucleon coupling and gives\cite{VVF}
\begin{equation}
f_\pi = 9.5 \pm 2.1_{exp} \pm 3.5_{th}  \times 10^{-7} .
\end{equation}
By comparison, nucleon pv experiments in $^{18}F$ are sensitive only to 
$f_\pi$ and give\cite{EGA}
\begin{equation}
f_\pi \leq 1.5 \times 10^{-7} .
\end{equation}
There appears to be a clear discrepancy between these two determinations of
$f_\pi$. However, the value determined from the anapole moment depends alo on 
the pv couplings of other mesons, primarily the 
$\rho$, to the nucleon via the combination $\sim f_\pi + \frac{1}{2} f_\rho$.
It could be that the $\rho$-N pv constant, determined from other pv 
experiments in light nuclei and the N-N system and consistent with each other 
and theory, is not correct; or perhaps some other effect has been omitted in 
the anapole calculations. The so-called "best value" of $f_\pi$, based on a 
quark model calculation is\cite{EGA}
\[f_\pi \simeq (4.6 \pm 4.6) \times 10^{-7} ,\]
where the error is often neglected. Nevertheless, there appears to be an 
a (experimental?) discrepancy which remains to be resolved. 

In addition to all of these studies, the electron has been used in studies of
time reversal invariance (TRI). Specifically, searches for an atomic electric
dipole moment in $^{199}Hg$ are sensitive to any simultaneous pv and TRI in
this system. E.N. Fortson et al.\cite{JPJ} obtain 
\begin{equation}
d_E (^{199}Hg) 
= - (1 \pm 2.4 \pm 3.6) 
\times 10^{-28} e-cm \; ,
\end{equation}
or 
\begin{equation}
\mid d_E(^{199}Hg)\mid  \leq 8.7 \times 10^{-28} e-cm 
\; (95\% \: confidence \: level) .
\end{equation}
This experiment has about the same sensitivity as measurements of the neutron
electric dipole moment and gives
\[ \theta \leq 10^{-10},\]
where  $\theta$ is defined by the term 
\begin{equation}
\cal L = - \theta \frac{\alpha_s}{8\pi} \epsilon_{\mu\nu\alpha\beta}G^{\mu\nu}
G^{\alpha\beta}, 
\end{equation} 
which would be the strong CP problem, except that $\theta \ll 1$.
But no one knows why $\theta$ should be so small.  At low energies, the pion
is the embodiment of QCD and the electric dipole moment comes about from a
T-odd $\pi$ - N coupling\cite{RJC,PH}
\[\cal L = - f_\pi^T  \bar{\psi} \vec{\tau} \psi \cdot \vec \phi\; ,\]
and  $f_\pi^T \leq 10^{11}$ from the upper limits of both the neutron and 
$^{199}Hg$ electric dipole moments.

I hope that I have convinced you that, despite its age, the electron and its 
interaction with nucleons and nuclei remains a very useful tool for testing 
fundamental interactions and learning important information about the 
structure of hadrons.

\end{document}